\RequirePackage{rotating}  
\documentclass[sigconf]{acmart}
\AtBeginDocument{%
  }

\setcopyright{acmlicensed}
\copyrightyear{2018}
\acmYear{2018}
\acmDOI{XXXXXXX.XXXXXXX}
\acmConference[Conference acronym 'XX]{Make sure to enter the correct
  conference title from your rights confirmation email}{June 03--05,
  2018}{Woodstock, NY}
\acmISBN{978-1-4503-XXXX-X/2018/06}

\begin{document}

\title{Interactive authoring of outcome-oriented lesson plans for immersive Virtual Reality training}

\author{Ananya Ipsita}
\affiliation{%
  \institution{Purdue University}
  \city{West Lafayette}
  \state{IN}
  \country{USA}
}
\email{aipsita@purdue.edu}

\author{Ramesh Kaki}
\affiliation{%
  \institution{Birla Institute of Technology and Science}
  \city{Pilani}
  \state{Hyderabad}
  \country{India}
}

\author{Mayank Patel}
\affiliation{%
  \institution{Purdue University}
  \city{West Lafayette}
  \state{IN}
  \country{USA}
}

\author{Asim Unmesh}
\affiliation{%
 \institution{Purdue University}
 \city{West Lafayette}
 \state{IN}
 \country{USA}
}

\author{Kylie A Peppler}
\affiliation{%
  \institution{University of California}
  \city{Irvine}
  \state{CA}
  \country{USA}
}

\author{Karthik Ramani}
\affiliation{%
  \institution{Purdue University}
  \city{West Lafayette}
  \state{IN}
  \country{USA}
}

\renewcommand{\shortauthors}{Ipsita et al.}

\begin{abstract}
  Immersive Virtual Reality (iVR) applications have shown immense potential for skill training and learning in manufacturing. However, authoring of such applications requires technical expertise, which makes it difficult for educators to author instructions targeted at desired learning outcomes. We present FlowTrainer, an LLM-assisted interactive system to allow educators to author lesson plans for their iVR instruction based on desired goals. The authoring workflow is supported by Backward design to align the planned lesson based on the desired outcomes. We implemented a welding use case and conducted a user study with welding experts to test the effectiveness of the system in authoring outcome-oriented lesson plans. The study results showed that the system allowed users to plan lesson plans based on desired outcomes while reducing the time and technical expertise required for the authoring process. We believe that such efforts can allow widespread adoption of iVR solutions in manufacturing training to meet the workforce demands in the industry.
\end{abstract}



\received{20 February 2007}
\received[revised]{12 March 2009}
\received[accepted]{5 June 2009}

\maketitle

\section{Introduction}
The U.S. manufacturing sector is facing a shortage of skilled workforce due to demographic shifts with an aging and retiring workforce and global shifts due to increased supply chain resilience \cite{nyt_manufacturing_good_old_days_2024, nyt_ai_aging_shift_2024, economist_altasia_2023}. These shifts call for the need to prepare students for a globally competitive environment. To meet these demands, training efforts must equip future professionals with both conventional and digital manufacturing skills \cite{nyt_ai_aging_shift_2024, economist_manufacturing_delusion_2023}. Immersive Virtual Reality (iVR) technology presents promising potential to enhance such efforts by providing immersive and interactive learning experiences without geographic constraints and at reduced cost, instructor time, and effort \cite{naranjo2020scoping}. Despite these benefits, the adoption of VR in manufacturing education is limited, which is partly due to the challenges faced by subject matter experts and educators to create content for training purposes \cite{badamasi2022drivers, scott2020investigation, jalo2021state, fernandez2017augmented, liagkou2019realizing}. This research provides an interactive system to reduce the barriers to authoring iVR content with an aim to facilitate iVR adoption for widespread skill training and learning.

Owing to the heavy dependence of technical modeling and programming expertise required for iVR content creation \cite{gaspar2020research, coelho2022authoring}, manufacturing educators typically collaborate with VR developers to author relevant content for their courses \cite{walczak2020semantic, gaspar2020research, cassola2021novel}. This collaboration, which typically happens on a contractual basis over a stipulated period, results in a packaged product at the end of the development cycle, with minimal room for variability in the final instruction design. Any demand for design modifications typically requires version change for the entire application, thereby limiting the scalability of instructions based on training requirements \cite{ketoma2023towards}. Moreover, this causes the subject matter experts (SMEs) to compromise with and plan the training programs around the rigid situation. To ensure that the iVR-based learning content is effective for achieving the desired goals, it is important that the content is tailored to the specific needs of different user groups and learning environments. Therefore, there is a need to propose workflows that empower SMEs with control over iVR instruction planning, catering to the varying needs of learners and the learning process.

Prior studies have provided workflows to author iVR content for skill training and learning. However, these works have either lacked focus on the reuse of content for varied requirements \cite{ chang2020exploring, yun2022immersive, tram2023intuitive, ipsita2024design, theofanidis2017varm, ipsita2021vrfromx} or highlighted the reuse aspect of authoring at an interaction level rather than learning level \cite{seo2001vr, zikas2020immersive}. The latter, while still useful, faces limitation to be used in educational context as they focus more on the technical aspects of the system design \cite{walczak2020semantic}. In contrast, our research provides an interactive workflow to authoring lesson plans for iVR-based instruction while still grounded in the pedagogical theories. In particular, the workflow utilizes the principles of Backward design which first identifies the desired learning outcomes and objectives, and then aligns the skills, assessment criteria and learning activities based on the intended learning outcomes \cite{mctighe2003backward, wiggins1998backward}. 

We present \textbf{\emph{FlowTrainer}}, an interactive workflow to enable authoring of lesson plans targeted towards desired learning outcomes. Using our workflow, SMEs can first author lesson plans using a web-based editor, and subsequently test the lesson plan in VR and iterate till the target lesson plan is achieved. The web-based editor provides a LLM (Large Language Model)-supported user interface where the user inputs can be guided interactively to generate a lesson plan by following a Backwards design approach. The process starts with defining the learning outcomes and then defining measurable objectives, skills and assessment criteria, and finally identifying the lesson plan in the form of a sequence of learning activities. In our work, the learning activities that act as the building blocks for the lesson plan are developed by the VR developers and stored in a library for educators to use based on the course requirements. These learning activities are categorized under the four instructional phases of $Introduction$, $Presentation$, $Practice$, and $Application$. These instruction phases provide a concrete framework for planning learning units \cite{wei2016opme, kruse2009gagne, merrill2002first, bybee2006bscs} and, when combined together, present a scalable framework which can be utilized to design a single lesson or a series of lessons, based on the target requirements of the training process. Using a welding use-case and based on our prior learning rationale extracted using Backward design \cite{ipsita2022towards}, a comparative user study was conducted with 8 experienced welders to test the usability and effectiveness of the system components to design the instructional flow for iVR-based learning units as compared to a baseline condition. The baseline condition is the preceding version of the current system which utilizes a similar workflow, but does not provide interaction capabilities to the user \cite{ipsita2024authoring}. From the user study, we aimed to answer the following research question.

\begin{itemize}
    \setlength\itemsep{0em}
    \item To what extent do the system components enable SMEs to flexibly author training scenarios based on varied learning objectives as compared to the baseline condition?
\end{itemize}

\noindent The user study results indicated that the current system enables SMEs to flexibly author training scenarios with ease and low mental workload under increasing complexity, supported by higher usability and reduced time of authoring as compared to the baseline condition. User feedback supported these findings by emphasizing the intuitive design, automatic task generation, and reduced effort in lesson creation. These characteristics make the current system suited for designing iVR training scenarios that must adapt to varied learning objectives, and thus positioning it as a scalable and cognitively accessible authoring tool for SMEs.

Thus, the contribution of our work are as follows:

\begin{itemize}
    \setlength\itemsep{0em}
    \item the system workflow design that enables SMEs to facilitate pedagogically driven planning of iVR-based learning units the Backward design approach \cite{mctighe2003backward, wiggins1998backward}, and
    \item the demonstration of the workflow design using a welding use case along with the evaluation and results from a comparative user study conducted with 8 experienced welders as compared to a baseline condition.
\end{itemize}

The article's structure is outlined as follows: In Section \ref{Section_2}, relevant literature is presented. Section \ref{Section_3} introduces key terminology to facilitate comprehension of the work. The design rationale is then detailed in Section \ref{Section_4}. Section \ref{Section_5} provides specific insights into the implementation details. The user study procedure is presented in Section \ref{Section_5} with the results described in subsection \ref{Section_6}. In Section \ref{Section_7}, the study results are discussed in greater detail. Finally, Section \ref{Section_8} presents the conclusion of the work.

\section{Related Work}\label{Section_2}
\subsection{iVR authoring in Education}
Authoring tools for content creation can solve the problems faced in the development of iVR-based training applications to some extent by making new VR content easier, faster, and more efficient \cite{coelho2022authoring, kaskalis2007multimedia}. However, the design of training scenarios remains particularly difficult due to the low level of code reuse, which often necessitates advanced programming skills. This obstacle can be overcome by reusing content, eliminating the need to develop the same asset repeatedly when creating or editing new experiences \cite{seo2001vr, zikas2020immersive}. Few of the above mentioned VR authoring tools allow reuse/import of content/assets enabling VR designers to significantly reduce their effort and generate custom user experiences that would otherwise be impossible. However, the high-level componentization approaches commonly employed in these content creation tools are heavily focused on technical aspects, resulting in a high level of complexity in the content design process \cite{walczak2020semantic}. This creates the requirement for knowledge engineering technicians to design training scenarios, and thus poses a challenge for SMEs to use these tools for VR training preparation. On the other hand, there has been some approaches of using machine learning methods to author curricula based on training needs. Kumar et. al. generated a customized curriculum for medical students based on deep learning techniques, utilizing data analysis and classification \cite{kumar2022customized}. The curriculum is generated using a gradient decision tree integrated with naïve Bayes, and learning approach recommendations are made using a fuzzy rules integrated knowledge-based recommendation system, with experimental results showing high accuracy and precision. However, such automated methods lack SME involvement in the content creation process, which limits their utility in the training process.

User-friendly content authoring tools are crucial for SMEs to design VR training scenarios based on their knowledge. These tools can play a significant role in reducing time and effort and promoting the use of VR in training \cite{walczak2020semantic}. Ziklas et. al. highlights the significance of innovative authoring platforms in enabling users with limited programming knowledge to easily design virtual environments \cite{zikas2023mages}. Cassola et. al. developed a method for creating semantic VR scenarios that can be used by users without advanced programming or 3D modeling skills \cite{cassola2021novel}. This approach empowers trainers and trainees to develop diverse and immersive learning scenarios for manufacturing and industrial applications, including certification. Coelho et. al. introduced intuitive interfaces for developers and content creators, providing real-time feedback on how the content will be presented to end-users \cite{coelho2019collaborative, coelho2021authoring}. This interface reduces the number of iterations required and potentially accelerates the development time of iVR applications. Wolfartsberger and Niedermayr proposed an authoring platform that empowers trainers to develop comprehensive courses within an immersive environment \cite{wolfartsberger2020authoring}. This platform enables trainers to specify elements, observe procedures, facilitate practice and testing, and assess trainees' performance. VR4Health functions as a training tool that allows teachers to monitor the learning process and provide feedback to students \cite{fairen2020vr4health}. It collects data on motivation, interest levels, and potential issues, facilitating the formulation of relevant questions for feedback sessions. Cassola et al. introduced a solution that enables SMEs to author specific sequence procedures for trainees, focusing on simulating immersive learning environments and increasing the availability of training components \cite{cassola2022design}. Lécuyer et al. provided a method to ease the scenario creation by enabling experts to record actions in VR to create action sequences, and then generate scenarios by utilizing those sequences \cite{lecuyer2020unveiling}. This approach aims to create engaging experiences for trainees. In terms of 3D content modeling for training scenarios, Walczak et al. proposed solutions that utilize domain knowledge representation techniques and the semantic web to describe content meaning in a standardized manner \cite{walczak2020semantic, lugrin2009alternative}. However, these tools have primarily been created as independent applications within specific contextual domains, overlooking pedagogically oriented approaches and neglecting their proper integration into the development cycle of iVR learning applications. Consequently, this limitation has led to a restricted adoption of such tools in pedagogical settings.

Unlike the development of iVR applications in the gaming and entertainment domains where it originated from, their effective use in professional settings requires incorporating expertise from various domains to design and implement effective training scenarios \cite{mishra2006technological, mahdi2018towards}. To accommodate the dynamic nature of diverse training needs, Mahdi et al. propose a stepwise process for developing pedagogically oriented VR training scenarios \cite{mahdi2018towards}. This process involves teachers expressing their needs, adapting the 3D environment, operationalizing scenarios, and conducting simulation/testing. They also introduce a custom-built tool for scenario development, although its limited features compared to game engines have hindered its widespread adoption. Saunier et al. suggest separating roles in the creation of VR training environments, identifying four main roles: designer, job expert, educational specialist, and teacher \cite{saunier2016methodology}. Each role actively contributes to the creation of pedagogically driven scenarios. Kavanagh et al. emphasize the use of VR learning by leveraging constructivist pedagogy and gamification in experience design \cite{kavanagh2017systematic}. Keskitalo highlights the pedagogical utilization of VR in education and training, particularly through simulation \cite{keskitalo2011teachers}. Similarly, Kemanji et al. emphasize the need for specifying and developing a dedicated pedagogical model for VR usage \cite{ketoma2023towards}. Building upon the aforementioned pedagogically driven research, our method presents a workflow design that is also driven by pedagogical models. It utilizes user-friendly components and enables SMEs to plan their instructional content based on varying training requirements, aiming to address the challenges faced in incorporating VR technology in educational settings.

\subsection{LLM-based lesson plan authoring in Education}

LLMs are increasingly used to support lesson planning and instructional design. \textit{LessonPlanner} helps novice teachers create pedagogy-aligned plans using Gagné’s Nine Events, improving quality and reducing workload \cite{fan2024lessonplanner}. Hu et al. guide LLMs to simulate teacher-student interactions and generate reflective feedback to iteratively enhance high school math plans \cite{hu2025exploring}. GPT-4 has demonstrated strong capabilities in setting objectives and organizing instructional content, though limitations remain in geometry and interdisciplinary modules \cite{hu2024teaching}. Zheng et al. introduced a three-stage process using Retrieval-Augmented Generation and self-critique prompting to generate high-quality lesson plans across elementary math topics \cite{zheng2024automatic}. GenAI tools with interactive mega-prompts enable tailored lesson plans based on classroom demographics and learning goals, leading to improved inclusivity and time savings \cite{karpouzis2024tailoring}. Additionally, fine-tuning techniques such as Chain-of-Thought prompting and LoRA have improved alignment between generated content and instructional frameworks like Gagné’s \cite{jia2025fine}. Our work builds on these approaches by introducing \textit{FlowTrainer}, which extends LLM-based lesson planning to immersive VR-based manufacturing training, supporting outcome-aligned instruction in technical domains.




\section{Clarification of Terms}\label{Section_3}

The following definitions will aid in the understanding of terminology used in this research work.
\begin{itemize}
    \setlength\itemsep{0em}
    \item \emph{Learning outcomes} indicate the relevant objectives (e.g., established goals, desired understandings) that will be addressed by the system. These will be useful in determining what students will know (e.g., what critical knowledge and skills will be acquired) and what students will be able to do as a result of the knowledge and skill acquisition \cite{adam2004using}.
    \item \emph{Learning objectives} are goals that target specific knowledge, skills, and dispositions taught in specific sections of a module \cite{nicholls2014key}. 
    \item \emph{Learning plan} is a description of how to design the learning experience and instructions to achieve the desired outcomes in learning \cite{nicholls2014key}.
    \item \emph{Learning activity} can be defined as a specific interaction of learner(s) using specific tools and resources, orientated towards specific outcomes \cite{grainne2007describing}.
    \item \emph{Learning module or unit} is an organized collection of teaching materials consisting of behavioral objectives, a sequence of learning activities, and provisions for evaluation \cite{robinson1972learning}.
\end{itemize}

\section{Design Rationale}\label{Section_4}

The process of instruction planning typically involves breaking down learning units or modules into segments of learning activities that revolve around relevant theories, concepts, or skills. Our proposed system for lesson planning of iVR-based instruction is based on the Backward design principles to effectively design outcome-oriented content \cite{mctighe2003backward, wiggins1998backward}. According to the Backward design model, the process begins with identifying the learning outcomes, and then subsequently working ``backwards'' to identify the learning objectives to achieve the desired goals. The next step is to identify the skills required to achieve the outcomes and objectives, and appropriate assessment strategies to measure the skills during the various learning stages. The final step is designing the learning activities required to meet the goal. Our system follows a similar approach to guide instructors to interactively provide these parameters to come up with the required lesson plan which is the sequence of learning activities.

The planning of a learning unit is further guided by the standardized instructional framework that groups learning activities into the four instructional phases of \emph{Introduction}, \emph{Presentation}, \emph{Practice}, and \emph{Application}. This instructional framework is common across various instructional design models and outlines the entire process of designing a learning experience, from audience analysis to implementation and revision \cite{wei2016opme, kruse2009gagne, merrill2002first, bybee2006bscs}. Furthermore, this instruction framework revolves around the learning outcomes, and prioritizes learner assessment and feedback throughout the process. Therefore, this can be easily integrated with the Backwards design framing of the curriculum \cite{wei2016opme}. We now briefly explain the four instructional phases utilized in the instruction planning process while providing examples of learning activities associated with each \cite{ConcordiaTeachingAcademy}. 

\begin{itemize}
    \setlength\itemsep{0em}
    \item \textbf{\emph{Introduction:}} This phase serves as the introduction to the topic. Some guiding principles for designing learning activities are: (1) enable students to remember prior experiences they can build on, (2) set expectations by sharing learning outcomes, and (3) provide necessary hook to arouse emotions and feelings in learners to facilitate better learning and retention. Examples include stories, videos, readings, diagnostic quizzes, etc.
    \item \textbf{\emph{Presentation:}} This phase guides students to engage with the content needed to achieve the learning objective. Some guiding principles for designing learning activities are: (1) determine what content is need-to-know vs. nice-to-know for achieving the objectives, (2) involve learners in actively learning the content within the situated learning context, and (3) make the content accessible and immersive for the learners. Examples include guided video tutorials, avatar demonstrations, narrations, and readings.
    \item \textbf{\emph{Practice:}} This phase enables students to perform practice activities targeting understanding of the specific content and performance of skills. Some guiding principles for designing learning activities are: (1) model and scaffold skills using necessary cues and prompts, (2) provide support mechanisms in terms of cues, hints and prompts, and (3) associate built-in evaluation and feedback. Examples include quizzes, Q\&A practice, and hands-on practical assignments.
    \item \textbf{\emph{Application:}} This phase allows students to apply new content and skills to authentic situations and integrating this with previous knowledge. Some guiding principles for designing learning activities are: (1) mimic real-world tasks in the field, (2) provide students opportunities to draw from their own diverse backgrounds and experiences, (3) associate built-in evaluation and feedback. Examples can include application tasks specific to the discipline.
\end{itemize}

Therefore, the design of learning units involves selecting appropriate activities from each instructional phase, taking into account the necessary support and practice learners will need to grasp a particular concept and achieve the associated learning outcome.

\subsection{\textbf{System Design and Implementation: }}\label{Section_4.1}
The system workflow design for planning iVR-based learning units comprises of three stages: (A) Developing a library of learning activities, (B) designing the instructional flow using a web-based editor, and (C) testing and verifying the instruction sequence of the learning activities. Before going into the details of each stage, we would like to mention that the current system is built on our previous work \cite{ipsita2024authoring}. The current system differs from the previous stage, particularly in Stage B where the web-based editor is provided with LLM-supported interactive capabilities with an aim to make the lesson planning easier for the user. The interaction is guided by following the principles of Backward design and thus aligns the learning units based on the target learning outcomes. Nevertheless, we would still go through the different stages to provide a comprehensive understanding to the user.

\begin{figure*}[h]
    \centering
    \includegraphics[width=\textwidth]{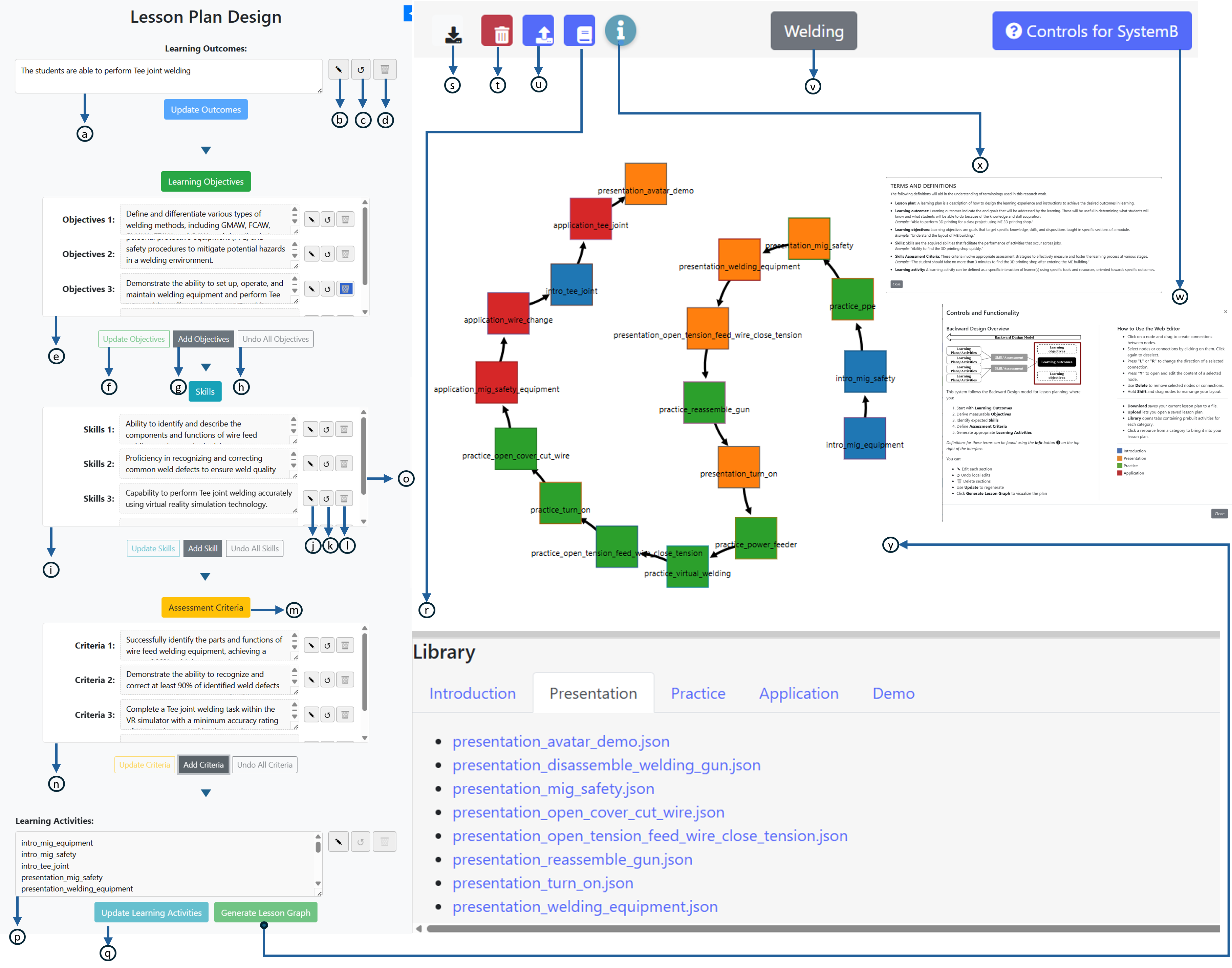}
    \caption{System Features of FlowTrainer: The left section of the UI presents an interface that assists user to generate a lesson plan by following the different stages of Backward Design. (a) The user can enter the learning outcomes, with features for (b) Edit, (c) Undo, and (d) Delete. The system generates (e) three learning objectives, (i) three skills and (n) three assessment criteria to measure the skills. Based on these a set of (p) learning activities are generated by the system. Local Edit, Undo and Delete features are used for individual objectives, skills or criteria (e.g. as shown in (j), (k), and (l)) where as there are global features such as add, delete and update objectives, skills and assessment criteria. The update of the hierarchical items happen in an hierarchical way, i.e., updating objectives have precedence over skills update however has lower precedence over updating outcomes. The learning activities can also be manually updated using (q) by adding or deleting learning activities. Collapsible buttons such as (m) and scroll bar such as (o) help in accessing the UI content. The lesson graph gets generated by using the button Generate Lesson graph which renders the lesson plan (graph sequence) on the web editor in the right section of the UI as shown in (y). The web editor has functionalities to edit the lesson graph using the resources from the library tab which can be opened by the Library button as shown in r. The user can click on any learning activities to bring the nodes to the web editor as required and create connections to add appropriately to the sequence. The user can save, delete or upload lesson plans using the buttons (s), (t), and (u) respectively. The details about the instructions on how to use the interface can be found in the modal instruction box which is accessed using the (w). Some definitions of the terminologies can be accessed using the modal instruction box (x). The button (v) at the top of the editor switches the mode between Demo and Welding for the user study.}
    \Description{System Features of FlowTrainer: The left section of the UI presents an interface that assists user to generate a lesson plan by following the different stages of Backward Design. (a) The user can enter the learning outcomes, with features for (b) Edit, (c) Undo, and (d) Delete. The system generates (e) three learning objectives, (i) three skills and (n) three assessment criteria to measure the skills. Based on these a set of (p) learning activities are generated by the system. Local Edit, Undo and Delete features are used for individual objectives, skills or criteria (e.g. as shown in (j), (k), and (l)) where as there are global features such as add, delete and update objectives, skills and assessment criteria. The update of the hierarchical items happen in an hierarchical way, i.e., updating objectives have precedence over skills update however has lower precedence over updating outcomes. The learning activities can also be manually updated using (q) by adding or deleting learning activities. Collapsible buttons such as (m) and scroll bar such as (o) help in accessing the UI content. The lesson graph gets generated by using the button Generate Lesson graph which renders the lesson plan (graph sequence) on the web editor in the right section of the UI as shown in (y). The web editor has functionalities to edit the lesson graph using the resources from the library tab which can be opened by the Library button as shown in r. The user can click on any learning activities to bring the nodes to the web editor as required and create connections to add appropriately to the sequence. The user can save, delete or upload lesson plans using the buttons (s), (t), and (u) respectively. The details about the instructions on how to use the interface can be found in the modal instruction box which is accessed using the (w). Some definitions of the terminologies can be accessed using the modal instruction box (x). The button (v) at the top of the editor switches the mode between Demo and Welding for the user study.}
    \label{fig:SystemDesign}
\end{figure*}

\textbf{\emph{Developing a library of learning activities:}}
The first step involves SMEs and VR developers working together to frame a set of learning activities by designing backwards from the overarching learning outcomes. Then, the development of the activities is performed by VR developers, while saving JSON configuration files for each. The configuration files contain the properties of the corresponding learning activities, that can be utilized by SMEs later to understand and/or edit the modifiable properties to personalize the training scenario. To test our approach, a library of 27 learning activities was developed for the welding use case, comprehensive enough to teach a class on basics of MIG welding including safety, welding equipment, welding wire change maintenance and Tee joint welding. The framing of the learning activities was based of our prior work where we collaborated with a team of SMEs, and by following the principles of Backward design from learning sciences came up with the learning activities that were sufficient to help learners gain the skills required to achieve the learning outcomes \cite{ipsita2022towards}.

\begin{figure*}
\centering
  \begin{sideways}
    \includegraphics[height=0.8\textwidth]{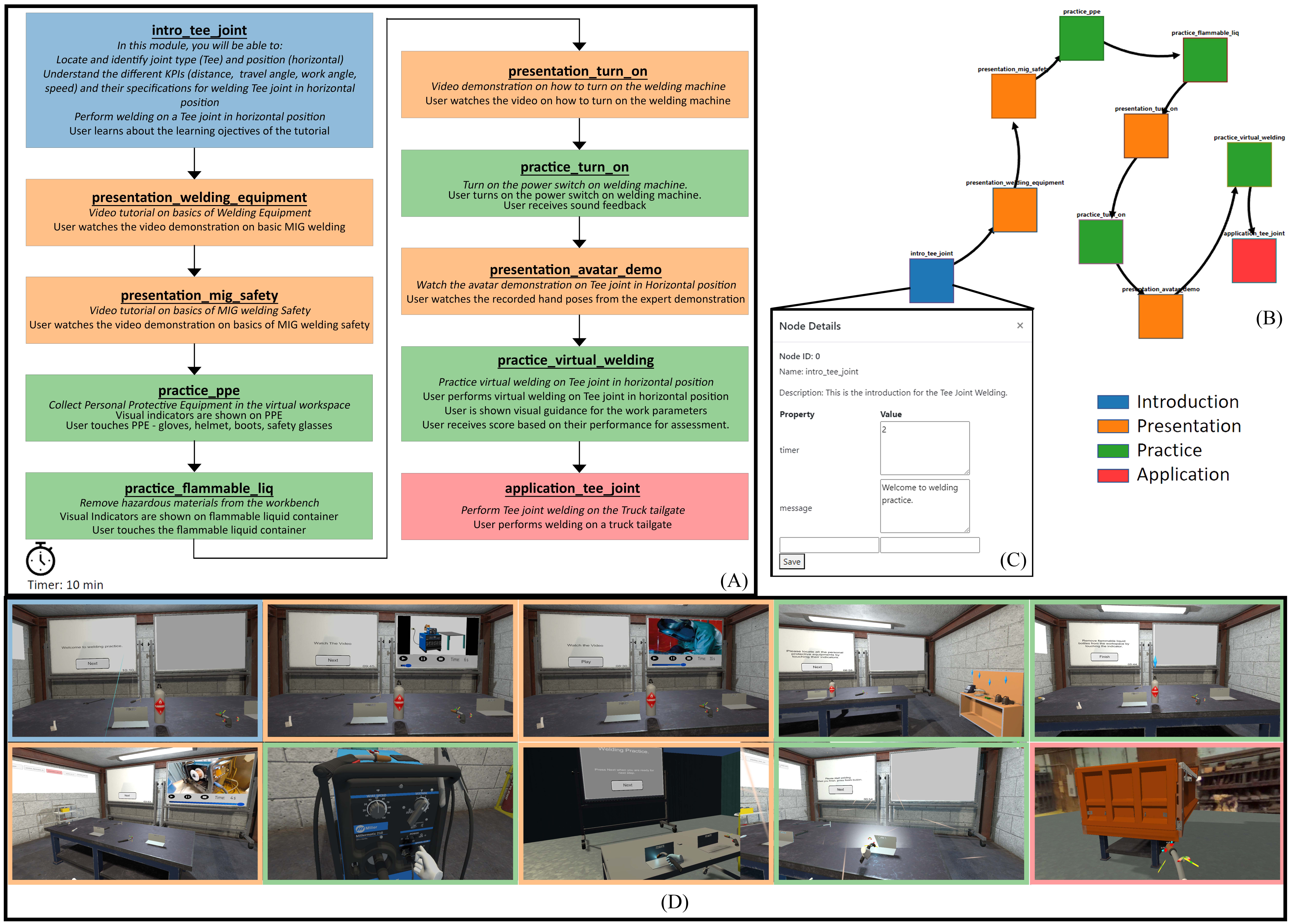}
  \end{sideways}
  \caption{An example implementation of the instructional planning of learning units using FlowTrainer: (A) Flowchart representation of the step by step planning of the learning unit, (B) Authoring the sequence of the iVR learning unit using the web-based editor, (C) The dialog prompt showing the details and editable properties of the node, and (D) Verification and testing of the learning unit in VR}
  \Description{An example implementation of the instructional planning of learning units using FlowTrainer: (A) Flowchart representation of the step by step planning of the learning unit, (B) Authoring the sequence of the iVR learning unit using the web-based editor, (C) The dialog prompt showing the details and editable properties of the node, and (D) Verification and testing of the learning unit in VR}
  \label{fig:FlowFigure}
\end{figure*}

\textbf{\emph{Designing the instructional flow using a graph-based flow editor:}}
To create the sequence of the learning activities catering to the needs of training, we developed a web-based flow editor as shown in Figure \ref{fig:SystemDesign}. The web-editor comprises of two parts: (A) the interactive user interface on the left section and (B) the graph-based editor on the right section. An example implementation is shown in Figure \ref{fig:FlowFigure}.

The nodes of the graph represent the learning activities which can appear on the interface by clicking corresponding learning activities from the library. SMEs can bring as many nodes onto the interface that they want to include in their training scenarios. Directed edge connections can be created between the nodes to create the instructional flow of the learning units. The interface has a resource library tab that allows users to view the available resources in the four specific categories of Introduction, Presentation, Practice and Application. The \emph{Controls} button is provided to show the detailed instructions on how to use the web-based editor. There are \emph{Upload}, \emph{Delete} and \emph{Download} buttons on the interface that allows the users to upload an existing scenario, delete the entire lesson plan on the interface and download the scenario flow from the interface respectively. Users can individually delete any node or edge on the web-based editor as well. There is a dialog box which appears on selecting a node and pressing key `Y' which shows the node details including any editable properties for the node. Using the editable properties, users can personalize the learning activities based on the configurable parameters available for the corresponding activity. In the current version of the work, the configurable parameters are limited to: (1) \emph{Timing:} This indicates the time duration to spend for the activity, (2) \emph{Message:} This indicates a short description shown to the learner, e.g., the objective of the learning activity. (3) \emph{Hint:} This indicates whether to show the audio and/or visual cues associated with the learning activity. The nodes are color-coded to help users distinguish between the node categories. Check prompts are also provided to check if any node in the graph contains more than one incoming or outgoing edges, or multiple sequences.

The user interface on the left guides users to interactively provide their requirements for the lesson plans. The process starts with providing the learning outcomes, then defining the objectives, skills, assessment criteria and finally the learning activities to achieve the outcomes. The interface is supported by LLM-based capabilities that can update the different parameters of the planning process based on the user's inputs and in a hierarchical manner. For example, updating the learning objectives updates the following parameters in the hierarchy, i.e., skills, assessment and learning activities. After the sequence of learning activities gets generated, the user can generate the lesson graph that renders the sequence on the graph-editor on the right. Users can iterate upon the lesson plan using the functionalities of the graph-based editor till the target requirements are achieved.

\textbf{\emph{Testing and verifying the instruction sequence of the learning activities:}}
Once the scenario flow is designed, SMEs can save the corresponding JSON file in designated file path. When the front end is started, the Unity application reads the JSON file from the path to dynamically generate buttons with the corresponding node names on the instruction screen. A Depth First Search (DFS) algorithm is utilized to generate the sequence for the learning activities to appear in the instruction screen in Unity. SMEs can either choose to view and verify the sequence in order. Or they can click on a specific button to jump to that particular activity. After verifying the sequence of the scenarios, users can go back and alter the lesson plan till they are satisfied with the scenario design.

\section{User Study}\label{Section_5}

\subsection{\textbf{Study Setup: }} The study was conducted with 8 participants (1 female, 7 male)(Age Range (number of users)=25-34 (8)) with prior welding experience who were recruited by word of mouth, email and flyer distribution. All users had prior experience with welding. 7 users had received prior training experience (informal or formal) in welding. Six users were novices in development for VR applications, two users rated themselves as intermediate in developing VR content. Five users had prior teaching experience, from which four of them designed curricula or learning instructions. Each user study was divided into three sessions where the task was to build six training scenarios, two per session using two systems SystemA and SystemB. SystemA is the graph-based editor without the interactive capabilities and is used as the baseline condition. SystemB is the FlowTrainer system with LLM-based capabilities. Each user was asked to plan the scenarios for three levels of task complexities in order using both systems. The three levels of task complexity were targeted towards gradually varying learning objectives. During this three-session study, the usability and mental workload of SystemB was evaluated while examining how the varying levels of task complexity affected the user experience and performance. Furthermore, the results were also compared with the baseline condition SystemA to gain insights on the user preferences and interaction. The study was approved under the IRB protocols.

\subsection{\textbf{Study Procedure: }} Users were first required to sign the consent forms and then proceeded to complete a short demographics survey. User demographics was collected about any prior experience with welding, VR, VR development, curriculum or instruction design, and teaching. Next, users were provided with the basic information in the form of a Word document and a PowerPoint presentation explaining the overall theme of the study. The information included the objective of the study, a short description about the four instructional phases, and the definitions of the stages involved in the Backward design lesson planning. To get accustomed to the system functionalities, users were asked to interact with the SystemA and SystemB using a cooking demo. For the cooking demo, they were asked to design a tutorial for pizza baking using the available learning activities under each phase that the users could use to plan the training scenarios. The available resources were enumerated in a table with their names and a short description, the sequence of which was randomly chosen. Users were asked to come up with a tutorial title, write the learning outcomes on the sheet, and then design the lesson plan using both systems, SystemA followed by SystemB.

After the demo, the actual study process for the welding simulator began. Users were asked to design lesson plans for the welding simulator. The available resources were enumerated in a table with their names and a short description, the sequence of which was randomly chosen. The PowerPoint presentation was prepared containing visual description of the available iVR-based learning activities, in the form of videos. After going through the resources, users were informed about the task of creating three training scenarios in three different sessions in order using SystemA and SystemB, the requirements of which were gradually disclosed immediately before the corresponding session started. Users were asked to consult the researcher present in the scene in case of any questions during the study. The order in which the user utilized the systems was counterbalanced to allow for within subjects comparision, while countering any order effects (e.g. learning, fatigue, etc.). The ordering resulted in the decision to include 8 users for the study. Each session had two tasks, during which the user performed the lesson planning using either SystemA or SystemB. Each lesson planning ended with a short survey to collect user ratings on the System Usability Scores (SUS) and workload (NASA-TLX) associated with the planning process using the system.

For the first session, users were first informed to imagine themselves to be the welding instructor of a class. The task was to plan a Tee joint Welding tutorial for a VR welding simulator to teach the imaginary class of students. Users were asked to come up with a tutorial title and the associated learning outcomes on the instruction document. This followed with the users designing the tutorial using the systems, one after another in the order pre-defined for the user. Each tutorial planning ended up with survey that collected scores for system usability and mental workload about the particular system.

During the second session, the users were informed about a change in requirements from the first session that the students in the imaginary class were novices in the field and did not have prior experience in welding. Considering this change in requirement, the new task was to build upon the previous tutorial plan to include additional information on basics of MIG welding, e.g., welding safety and equipment. The study procedure remained the same as before where users modified the previous tutorial plan based on the new information using both systems.

During the third session, the requirement changed again. It was realized that the students were needed to be taught about the welding wire change maintenance, and thus the tutorial had to be modified based on the updated requirements. Same as before, the users were asked to modify the tutorials using both systems. After the three sessions were finished, the users were asked to finish a survey questionnaire asking them to record their responses about their preferences with using both systems. The survey also asked subjective responses regarding what features they liked or disliked in both systems and suggestions for improvement. The study lasted for approximately 1.5 hours and each user was given \$15 compensation.

Data was collected for each user in terms of: (i) time required to finish each session task, (ii) JSON files for the training scenarios, (iii) 5-point Likert scale-based custom questionnaire to evaluate the usability, user perception and experience about using the system components, and (iv) subjective feedback. The JSON files from the training scenarios were analyzed to collect the sequence length distributions across the different sessions. A researcher recorded the duration of the tasks in real time during the three sessions. Each user study session was recorded by using a screen capture, in case a review was needed or timing was missed. Data about the welding training scenarios was recorded from the flow editor for the three sessions. This data was analyzed later to obtain information about the ability of the system to enable flexible authoring of training scenarios based on variable learning objectives. After each task, users recorded their experience with the system using a 5-point Likert survey. The subjective comments and suggestions from the users during the final survey were later used in the paper to explain the study results and inspire future design insights.

\subsection{\textbf{Results: }}\label{Section_6} The quantitative results of the study provide a comprehensive comparison of SystemA and SystemB across usability, workload, and performance metrics in terms of timing and sequence length. After checking that the assumptions of normality (Shapiro-Wilk test, $p>$0.05) were not met, non-parametric methods were utilized to find any statistical significance in the results. The survey results are described while providing the scores for average (Avg) and standard deviation (SD).

\begin{figure*}
    \centering
    \includegraphics[width=\textwidth]{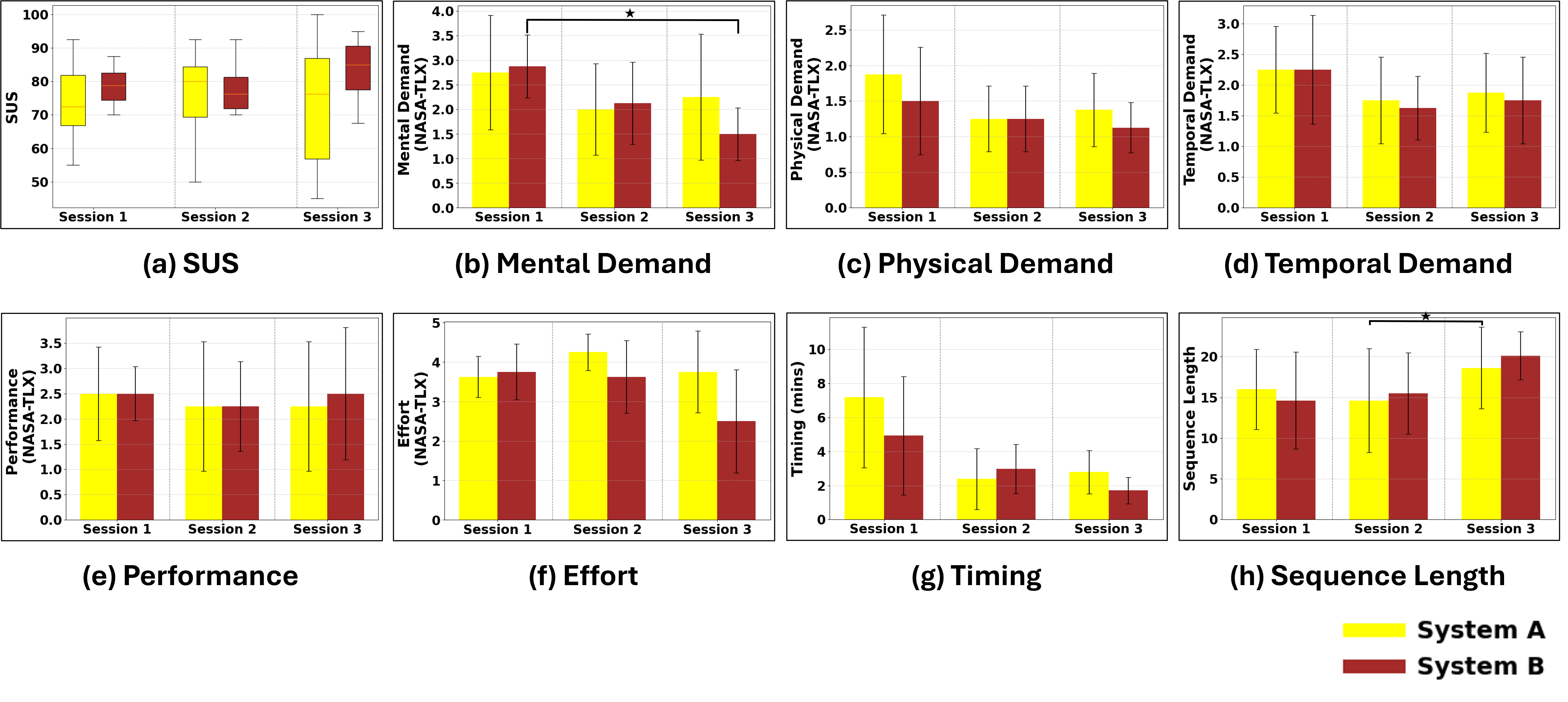}
    \caption{Quantitative Results from the Survey Questionnaire showing the comparision between SystemA (web-based editor without LLM capabilities) and SystemB (web-based editor with LLM capabilities): (a) System Usability Scores (SUS) across the three sessions, (b-f) NASA-TLX ratings for Mental Demand, Physical Demand, Temporal Demand, Performance and Effort across the three sessions, (g) Time of Completion of tasks across the three sessions, and (h) Sequence Length Distribution across the three sessions.}
    \Description{Quantitative Results from the Survey Questionnaire showing the comparision between SystemA (web-based editor without LLM capabilities) and SystemB (web-based editor with LLM capabilities): (a) System Usability Scores (SUS) across the three sessions, (b-f) NASA-TLX ratings for Mental Demand, Physical Demand, Temporal Demand, Performance and Effort across the three sessions, (g) Time of Completion of tasks across the three sessions, and (h) Sequence Length Distribution across the three sessions}
    \label{fig:SurveyResults}
\end{figure*}

\textbf{\emph{System Usability Scores for SystemA and SystemB across sessions:}}
System Usability Scale (SUS) scores were computed for each session, with SystemB showing higher usability ratings than SystemA during Sessions 1 and 3. In Session 1, SystemB had an average SUS score of 74.38 (SD = 8.53) compared to SystemA with an average of 72.19 (SD = 12.35). In Session 2, SystemB scored 75.00 (SD = 13.36) while SystemA slightly surpassed it with 76.56 (SD = 13.82). However, in Session 3, SystemB had a significant advantage with a score of 83.12 (SD = 10.42), compared to SystemA with an average of 73.44 (SD = 19.64).

Statistical analysis using the Friedman test showed that task complexity did not significantly impact usability ratings for either system. For SystemA, the test yielded $\chi$² = 1.31, $p =$ 0.5194, and for SystemB, $\chi$² = 4.22, $p =$ 0.1211. Post-hoc Wilcoxon tests also showed no significant pairwise differences across complexity levels for either system after Bonferroni correction. For SystemA, all comparisons (e.g., Session1 vs. Session2, Session2 vs. Session3) had p-values > 0.52, and for SystemB, while some comparisons approached significance (e.g., Session1 vs. Session3 at $p =$ 0.093), none met the corrected threshold ($p =$ 0.0167). The box plots showing the SUS scores can be found in Figure \ref{fig:SurveyResults}(a).

\textbf{\emph{NASA-TLX ratings for SystemA and SystemB across Sessions:}}
The NASA-TLX results across the three sessions reveal how SystemA and SystemB compare in terms of perceived workload. The results can be found in can be found in Figure \ref{fig:SurveyResults}(b-f).

In Session 1, both systems exhibited comparable mental workload, with SystemA averaging 2.75 (SD=3), and SystemB slightly higher at 2.88 (SD=3), though variability was lower in SystemB. Physical and temporal demands were low overall, while effort was slightly greater for System with an average of 3.75 (SD=4), compared to SystemA with an average of 3.63 (SD=4). Performance ratings remained equal with an average of 2.5 (SD=3), indicating similar perceived effectiveness. In Session 2, SystemA had slightly lower mental demand (Avg=2.00, SD=2) than SystemB (Avg=2.13, SD=2), but required more effort in SystemA (Avg=4.25, SD=4) as compared to SystemB (Avg=3.63, SD=4). Physical and temporal demands were identical across systems. In Session 3, SystemB outperformed SystemA, with lower averages across all workload dimensions, mental demand dropped in SystemB (Avg=1.5, SD=2) as compared to SystemA (Avg=2.25, SD=2), and effort was significantly lower in SystemB (Avg=2.5, SD=3) as compared to SystemA (Avg=3.75, SD=4). These values indicate that as task complexity increased, SystemB was perceived as less mentally and physically demanding.

Statistical analyses further confirm these observations. Friedman tests showed that task complexity had a significant effect on mental workload in SystemB ($\chi$² = 11.27, $p =$ 0.0036), but not in SystemA ($p =$ 0.0578). Post-hoc Wilcoxon tests revealed a significant difference in SystemB between Session1 and Session3 ($p =$ 0.0158), whereas SystemA showed no significant pairwise differences. For physical workload, the effect of task complexity was significant in SystemA ($\chi$² = 6.13, $p =$ 0.0468) but not in SystemB ($p =$ 0.1738), although post-hoc tests revealed no significant pairwise differences in either system after Bonferroni correction. Neither system showed significant effects of task complexity on temporal demand, performance, or effort based on Friedman test results (all $p =$ 0.05). Furthermore, all post-hoc comparisons across these metrics failed to reach significance, reinforcing the conclusion that only mental workload in SystemB and physical workload in SystemA were impacted by increasing task complexity.

These findings demonstrate that SystemB maintains lower and more consistent workload under complexity, particularly in terms of cognitive load. While SystemA was more sensitive to task complexity in physical demand, it did not lead to meaningful differences in user performance or overall effort. This supports the suitability of SysytemB for flexible, scalable use in authoring VR training content, especially for more cognitively demanding scenarios.

\textbf{\emph{Time of Task Completion for SystemA and SystemB across sessions:}} In terms of task completion time, SystemA showed a noticeable drop from Session 1 to Sessions 2 and 3, with average timing decreasing from 7.19 seconds (SD = 4.13) to 2.39 (SD = 1.79) and 2.79 seconds (SD = 1.28), respectively. SystemB also showed a decrease, starting from 4.94 seconds (SD = 3.48) in Session 1 and dropping to 2.98 (SD = 1.44) in Session 2 and 1.71 seconds (SD = 0.77) in Session 3. These results suggest that users required less time as they became more familiar with the systems or as the task types changed. However, Friedman tests revealed no statistically significant effect of task complexity on timing for either system (SystemA:$\chi$² = 4.75, $p =$ 0.0930; SystemB:$\chi$² = 5.25, $p =$ 0.0724), and the Wilcoxon test showed no significant difference in timing variability between the systems ($p =$ 0.3828). Post-hoc tests further confirmed that none of the pairwise comparisons across task complexities reached significance under Bonferroni correction. The results can be found in can be found in Figure \ref{fig:SurveyResults}(g).

\textbf{\emph{Sequence Length Distributions for SystemA and SystemB across Sessions:}} Regarding sequence length, which reflects the number of learning steps or nodes used in the lesson plans, SystemA increased from 16.00 (SD = 4.93) in Session 1 to 18.62 (SD = 5.01) in Session 3, while SystemB increased from 14.62 (SD = 5.95) to 20.12 (SD = 2.95). Friedman tests revealed a significant effect of task complexity on sequence length for both systems (SystemA:$\chi$² = 10.57, $p =$ 0.0051; SystemB:$\chi$² = 6.07, $p =$ 0.0482). However, the Wilcoxon test found no significant difference between systems regarding how much complexity affected sequence length ($p =$ 0.4990). In post-hoc analysis, only SystemA showed a significant difference between moderate and complex tasks ($p =$ 0.0078. In SystemB, no pairwise comparison reached significance under Bonferroni correction. Additionally, the Wilcoxon test comparing variability in sequence length between systems found no significant difference ($p =$ 1.000). The results can be found in can be found in Figure \ref{fig:SurveyResults}(h).

\textbf{\emph{Custom questionnaire on User Experience:}} From the final survey questionnaire on a 1-5 Likert Scale \emph{[1=Strongly Disagree, 5=Strongly Agree]}, it was observed that participants generally leaned toward neutral responses on most statements comparing SystemA to SystemB. They somewhat disagreed that SystemA was simpler to use (Avg=2.43, SD=1.05) and easier to get familiar with (Avg=2.57, SD=1.4). Opinions were close to neutral regarding whether SystemA’s interface better captured their intent (Avg=2.71, SD=1.03) and whether it helped with tutorial planning (Avg=2.43, SD=1.18). Users slightly agreed that SystemA allowed easier modification of tutorials based on changing requirements (Avg=3.43, SD=1.29) and iteration on existing plans (Avg=3.71, SD=1.16). They somewhat disagreed that SystemA required a lot of preparation (Avg=2.43, SD=1.18). Ratings were neutral on SystemA’s usefulness for creating VR lessons (Avg=3.00, SD=1.51) and slightly below neutral for instructors using it in classrooms (Avg=2.71, SD=1.03).

\section{Discussion}\label{Section_7}

Based on the results and subjective feedback from users, we now try to answer the research question that we posed before. \textbf{\emph{To what extent do the system components enable SMEs to flexibly author training scenarios based on varied learning objectives as compared to the baseline condition?}}

Based on the quantitative results, it is observed that SystemB demonstrates a stronger capacity to enable SMEs to flexibly author training scenarios aligned with varied learning objectives when compared to SystemA. Going further, we discuss this in more detail while providing qualitative responses from users during the survey.

From the usability and cognitive load analysis, SystemB received consistently higher and more stable SUS scores especially in Sessions 1 and 3. This suggests that it provides a more user-friendly interface for content creation. The significant jump in usability scores during Session 3 (SystemB: 83.12 vs. SystemA: 73.44) suggests that SystemB scales better with the cognitive demands of complex tasks. As pointed out by a user, ``\emph{I can give a prompt, and it will make all the required steps, which if I feel needs to be changed or I might have missed, I can do that. SystemA is more time-consuming.}'' Another user mentioned, ``\emph{It was easier to use, and I did less effort in typing out all the subsections of the lesson plan. It was better not to interact with the floating bubbles like in SystemA.}''

NASA-TLX results further support this conclusion. Mental workload in SystemB significantly decreased with increasing task complexity. This indicates that users found it cognitively easier to build complex training scenarios. On the other hand, increasing physical demand with complexity in SystemA may indicate limitations in its authoring affordances. As pointed out by one user, ``\emph{As the nodes grew and the arrows started crossing each other while editing, it was hard to keep track of everything.}'' The need to repeatedly add and connect nodes manually even for straightforward scenarios, was mentioned as a drawback. One user suggested, ``\emph{simplify the input and edit mechanism to reduce all the bubbles with so many crossings to avoid confusion.}''

The trends for time of task completion showed that both systems improved with user familiarity. However, SystemB showed more efficient performance for complex tasks, with the shortest completion time of 1.71 minutes in Session 3, compared to 2.79 minutes in SystemA. This suggests that SystemB has a smoother learning curve. As pointed out by an user, ``\emph{It is easy to make the entire class presentation with a command and easy to modify.}''

In terms of authoring capacity as reflected by sequence length, both systems showed increased sequence lengths with task complexity. This suggests more detailed content creation over time. However, the average sequence length in Session 3 (20.12) for SystemB was higher than that of SystemA (18.62). While only SystemA showed a statistically significant jump between moderate and complex tasks, this might reflect less nuanced control over complexity. As one user put it, ``\emph{If there was a way to change the order of skills learning and practicing, then SystemB would be better than A.}''

Responses from the custom questionnaire showed that, although general sentiment remained neutral, users leaned slightly in favor of SystemB for usability and ease of learning. However, SystemA was occasionally preferred for modifying and iterating on content, potentially because it offers a more manual, hands-on workflow. One participant mentioned, ``\emph{The fact that I can choose the order in which I would like to plan the tutorials}'' as a favorable aspect of SystemA, while also noting that ``\emph{It was cumbersome to know all the resources in A to plan the tutorials.}'' For SystemB, users appreciated features such as ``\emph{automatic generation of tasks}'' and ``\emph{lesson plan generation, setting the learning outcomes,}'' though a few pointed out limitations, such as ``\emph{manual typing can be a bit hectic}'' and ``\emph{the generated results looked similar in most cases.}''

Although the small sample size and the lack of experience in teaching and instruction for some users may affect the results, we have considered the counterbalancing effects for the comparative study. We acknowledge that this is a limitation in the study and the results may improve from a more targeted population and increased sample size. Nevertheless, the results presented in this work are still useful and can guide researchers in this field to facilitate the use of VR for learning purposes.

\section{Conclusion}\label{Section_8}
We presented FlowTrainer, a system workflow design that enables subject matter experts to plan of iVR-based learning units based on the training requirements. The workflow begins with VR developers developing a library of learning activities with configurable parameters, based on the learning outcomes and assessment identified using the Backward design approach. These learning activities are grouped under the four instructional phases of Introduction, Presentation, Practice and Application, based on the scalable instructional framework for instruction design. Using the learning activities as the building blocks, subject matter experts can plan their learning units using a graph-based editor. These units can be later verified in VR to test their effectiveness in meeting desired learning objectives. Using a welding use case, our user study with 8 experienced welders showed the effectiveness of the system to author flexible training scenarios based on varied learning objectives. The positive ratings from the users indicated usability of such workflows in designing iVR-based learning units. We believe that the insights obtained from the research can be helpful towards enhancing the adoption of the immersive technology in educational settings.

\begin{acks}
We thank the reviewers for their invaluable feedback. Any opinions, findings, and conclusions or recommendations expressed in this material are those of the authors and do not necessarily reflect the views of the funding agency.
\end{acks}

\bibliographystyle{ACM-Reference-Format}
\bibliography{sample-base}

\end{document}